\documentclass[aps,prl,reprint,superscriptaddress,amsmath,amssymb,showpacs,floatfix]{revtex4-1}
\usepackage{graphicx}
\usepackage{bm}
\usepackage{color}

\begin{document}

\newcommand{\be}{\begin{equation}}
\newcommand{\ee}{\end{equation}}

\title{Epidemic Threshold of Susceptible-Infected-Susceptible Model on Complex Networks}

\author{Hyun Keun Lee}
\affiliation{Department of Physics, University of Seoul, Seoul 130-743,
Korea}
\author{Pyoung-Seop Shim}
\affiliation{Department of Physics, University of Seoul, Seoul 130-743,
Korea}
\author{Jae Dong Noh}
\affiliation{Department of Physics, University of Seoul, Seoul 130-743,
Korea}
\affiliation{School of Physics, Korea Institute for Advanced Study,
Seoul 130-722, Korea}

\date{\today}

\begin{abstract}
We demonstrate that the susceptible-infected-susceptible~(SIS) model on complex
networks can have an inactive Griffiths phase characterized by a slow
relaxation dynamics.
It contrasts with the mean field theoretical prediction that the SIS model
on complex networks is active at any nonzero
infection rate.
The dynamic fluctuation of infected nodes, ignored in the mean field
approach, is responsible for the inactive phase.
It is proposed that the question whether the epidemic threshold of the
SIS model on complex networks is zero or not
can be resolved by the percolation threshold in a model where nodes are
occupied in the degree-descending order.
Our arguments are supported by the numerical studies on scale-free network
models.
\end{abstract}


\maketitle
Epidemic spreading is a common phenomenon in networked systems.
Diseases spread from individual to individual through a contact network
and computer viruses spread through the Internet.
Since it has a huge impact on stability,
epidemic spreading on complex networks has been attracting a lot of interests
during the last decades~\cite{Barrat08}.
Those studies have focused on both theoretical issues such as nonequilibrium critical phenomena~\cite{Dorogovtsev08}
and practical issues such as searching for
an efficient immunization strategy~\cite{Cohen03,Kitsak10}.
The SIS model is a paradigmatic
epidemic spreading model where an infected individual becomes susceptible (or healthy) at a unit rate
and infects its susceptible neighbor at a rate $\lambda$.
We consider the SIS model on complex networks whose
degree distribution $P(k)$ denoting the fraction of nodes with degree $k$ is
broad~\cite{Albert02}.

Pastor-Satorras and Vespignani proposed a so-called heterogeneous
mean-field~(HMF) theory
for complex networks~\cite{Pastor-Satorras01a}.
According to it,
the epidemic threshold of the SIS model, above
which the system is in an active phase with a finite density of infected nodes,
is given by $\lambda_{\rm c} = \langle k \rangle / \langle k^2 \rangle$ with
$\langle k^n\rangle = \int dk k^n P(k)$.
Specifically, in scale-free~(SF) networks
characterized by $P(k) \sim k^{-\gamma}$ with a degree
distribution exponent $\gamma$~\cite{Albert02},
$\lambda_{\rm c} = 0$ for $\gamma\leq 3$
while
$\lambda_{\rm c} > 0$ otherwise~\cite{Pastor-Satorras01b}.
The HMF theory, which becomes exact in the annealed network
limit~\cite{Castellano08,Noh09,Lee09}, turns out to be useful
in studying various physical problems on complex
networks~\cite{Dorogovtsev08}.

Meanwhile, a recent mean filed (MF) study~\cite{Castellano10} on
rather realistic {\it quenched} networks reports that the epidemic
threshold vanishes~($\lambda_{\rm c}=0$) in any network with diverging maximum degree.
It implies that an epidemic spreading cannot
be prevented on complex networks with an unbounded degree
distribution. This study attracts much interest and is followed by a series
of works~\cite{Mieghem12,Castellano12,Goltsev12,Ferreira12}.
The MF theory on quenched networks will be referred to as the
quenched mean-field~(QMF) theory.
The discrepancy between the two MF theories makes it urgent
to study the SIS model beyond a MF level.

In this Letter, we challenge the MF approach by taking account of
the dynamic fluctuation of infected nodes. This effect turns out to be
crucial in determining whether the epidemic threshold of the SIS model
on complex networks is zero or not. We find that the active phase
predicted by the QMF theory near $\lambda = 0$ actually corresponds to
the Griffiths phase~\cite{Hooyberghs03,Vojta06,Munoz10}
where the density of the infected nodes decays to zero more slowly
than an exponential decay, unless the active nodes in the QMF theory
form a percolating cluster.
It is proposed that zero/nonzero epidemic threshold of the SIS model
is inherited from zero/nonzero percolation threshold in a model
where nodes are occupied in the degree-descending order.
Such a specific percolation
will be referred to as the degree-ordered percolation (DOP).
Our argument is confirmed in the numerical studies on the $(u,v)$-flower
model~\cite{Rozenfeld07} for scale-free networks.
We finally apply the DOP to survey whether $\lambda_{\rm c}$ of the SIS model
should be zero or not on random SF networks, which remains unsettled
in model simulations due to a strong finite-size effect~\cite{Ferreira12}.

We begin with a review on the QMF theory for the SIS model.
Let $\rho_i(t)$ be the infection probability of node $i$ at time $t$.
Then the rate equation reads
\be\label{SRE}
\frac{d\rho_i(t)}{dt} = -\rho_i(t)+(1-\rho_i(t))\sum_j a_{ij} \lambda \rho_j(t)~,
\ee
where $a_{ij}$ is an element of the adjacency matrix assigned with $1$ if there is an edge between nodes $i$ and $j$ or $0$ otherwise.
The first term of the R.H.S. of Eq.~(\ref{SRE}) is the recovery rate
reducing the infection probability and the second term is the infection rate
given by the product of the susceptible probability and
the infection trial rate by
infected neighbors.

The QMF approach focuses on the linear stability analysis of
the zero fixed point~($\rho_i(0)=0$ for all $i$) of Eq.~(\ref{SRE}), which
corresponds to a configuration
of the inactive phase.
It is easy to show that the fixed point becomes unstable
as soon as $\lambda \Lambda_1 > 1$ for the largest eigenvalue
$\Lambda_1$ of $\{a_{ij}\}$.
This leads to the conclusion
$\lambda_{\rm c}^{\rm QMF}=1/\Lambda_1$
for the epidemic threshold of the QMF theory~\cite{Wangetal}.
Although appealing, it has some controversial points.
Most of all, it predicts $\lambda_{\rm c} = 0$
in any network with the diverging maximum degree. In an arbitrary graph
with the maximum degree $k_{\rm max}$, the largest eigengenvalue satisfies
an inequality
$\sqrt{k_{\rm max}}\leq \Lambda_1 \leq k_{\rm max}$~\cite{Stevanovic03}.
This gives $\lambda_{\rm c}^{\rm QMF} = 0$ in the
$k_{\rm max}\rightarrow \infty$ limit.
An alternative interpretation of $\lambda_{\rm c}^{\rm QMF}$ follows recently
in Ref.~\cite{Goltsev12}, which claims that a property of the eigenvector
corresponding to $\Lambda_1$ plays an important role in epidemic prevalence.

As a counterexample to the QMF conclusion,
it is instructive to consider a star graph consisting of a hub at center
and $k_{\rm max}$ linear chains of length $L$ emanating from it.
The total number of nodes is $N = k_{\rm max} L+1$.
It is straightforward to show that $\Lambda_1 \rightarrow k_{\rm max}
/ \sqrt{k_{\rm max}-1}$ for large $L$. Hence, $\lambda_{\rm c}^{\rm QMF} = 0$
in the infinite $k_{\rm max}$ limit.
On the other hand, the steady state solution of
Eq.~(\ref{SRE}) is given by
$\rho_r \propto [2\lambda/(1+\sqrt{1-4\lambda^2})]^r$ for
$1/\Lambda_1 < \lambda \le 1/2$ and $\rho_r = (2\lambda-1)/(2\lambda)$
for $\lambda \gtrsim 1/2$,
where $1\ll r \ll L$ denotes the distance from the hub.
So the epidemic order parameter $\rho \equiv \lim_{L,k_{\rm max}\rightarrow \infty} \sum_i \rho_i/N$ is given by
\be\label{pt}
\rho = \left\{
\begin{array}{ll}
0 &~{\rm for}~\lambda \le 1/2 \\ [2mm]
\lambda-1/2 &~{\rm for}~\lambda \gtrsim 1/2 \ .
\end{array} \right.
\ee
Namely, $\lambda_{\rm c} = 1/2$.
This example demonstrates that the linear stability analysis against
the inactive state alone is not sufficient in determining the
threshold of the SIS model on complex networks.

In order to overcome the shortcoming of the previous QMF approach, we
suggest to take account of the other eigenvectors of the adjacency matrix
than the leading one with the largest eigenvalue $\Lambda_1$. This approach
was also taken in Ref.~\cite{Goltsev12}.
More importantly, we also suggest to take account of
the fluctuation of active nodes due to the stochasticity
in updating each node's state, which is ignored in most
MF approaches.

For convenience,
we label the nodes in the degree-descending order:
$k_1 \geq k_2 \geq \cdots \geq k_N$.
Recent studies show that the $n$-th largest eigenvalue of the adjacent matrix
of a random network is $\Lambda_n \sim \sqrt{k_n}$ for large
$k_n$~\cite{Chung03} and the corresponding eigenvector
is localized around the node $n$~\cite{Nadakuditi12}.
These findings imply that the steady state solution of Eq.~(\ref{SRE}) at
small values of $\lambda$ consists of local active domains around
high degree nodes: Each high degree node $n$ behaves like an independent
local hub with its own activation threshold given by
$\lambda_n = 1/\Lambda_n \sim 1/\sqrt{k_n}$ and the size of a local active
domain is given by $\sim \lambda k_n$.
Independence of the local active domains in the small $\lambda$ limit
is guaranteed only when higher degree nodes are distant enough
from each other~(this will be clarified later).
A network with such a property will be referred to as
the {\it unclustered} network.
For $1/\Lambda_n < \lambda < 1/\Lambda_{n+1}$, there appear the local active
domains of size $\sim \lambda k_i$ around all nodes $i\le n$.
Thus one may expect
\be\label{qbeta}
\rho \sim
\int_{1/\lambda^2}^{\infty} dk\ (\lambda k)P(k)~,
\ee
which yields $\rho \sim \lambda^{2\gamma - 3}$ for a random SF network with a degree exponent $\gamma$.
Note that the refined QMF theory still predicts that $\lambda_c = 0$ with the
order parameter exponent $\beta = 2\gamma-3$.

\begin{figure}
\includegraphics*[width=\columnwidth]{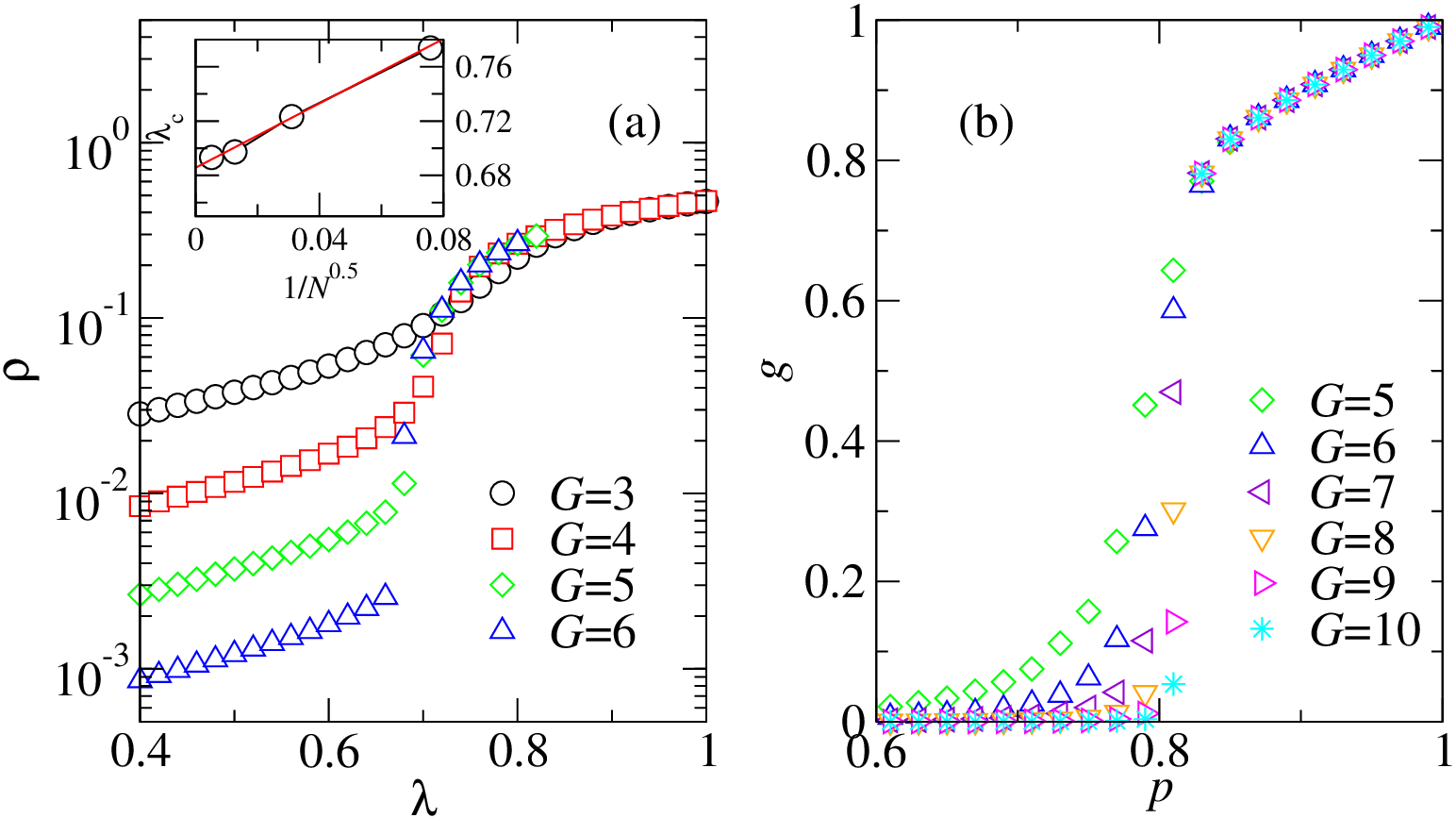}
\caption{(Color online) Density of infected nodes in the $(3,3)$-flowers in
(a). The largest cluster density in the DOP (see text) on the same flowers
as in (b).}
\label{fig1}
\end{figure}

A numerical evidence, however, shows Eq.~(\ref{qbeta}) is not valid.
We have performed Monte Carlo simulations for the SIS model on
SF networks generated by the $(u,v)$-flower model~\cite{Rozenfeld07}.
It is a deterministic hierarchical model: One starts from two nodes
connected with an edge~(zeroth generation). Then, every link in a $G$-th
generation is replaced with the two $u$- and $v$-link-long paths in the
next $(G+1)$-th generation.
It results in a SF network with $\gamma = 1+\ln(u+v)/\ln2$ in the
$G\rightarrow \infty$ limit~\cite{Rozenfeld07}.
The $(u,v)$-flower model is particularly useful because one can generate an
unclustered network easily. If $u>1$ and $v>1$, the degree of all
nodes is doubled and the distance between them becomes farther
after
each iteration.
We used the $(3,3)$-flower model in simulations which certainly belongs to
the unclustered network.
The numerical data shown in Fig.~\ref{fig1} (a) strongly suggest a transition
at nonzero $\lambda_{\rm c}$. The threshold can be estimated from the
peak positions of the susceptibility $\chi \equiv N (\langle \rho^2\rangle -
\langle \rho\rangle^2)/\langle \rho\rangle$ where $N$ is the total number of
nodes~\cite{Ferreira12}.
The inset in Fig.~\ref{fig1} shows the peak position is extrapolated to
$\lambda_{\rm c} \simeq 0.65(5)$ in the infinite system size limit.

The origin of the inconsistency is an irreversible fluctuation
within local active domains due to the stochasticity in microscopic dynamics.
Consider a local domain consisting of $V$ nodes where each node is
in the infected state with a probability $r$ and the susceptible state
with a probability $(1-r$). Then, no matter how rare, there exists a moment
when all nodes recover simultaneously by chance.
It takes place after a characteristic time
$\tau_V \sim (1-r)^{-V}\sim e^{r V}$.
Once being recovered, the local domain never returns to an active phase
unless externally activated.
Considering this effect of the irreversible fluctuation,
Eq.~(\ref{qbeta}) should be replaced by
\be\label{relax_dyn}
\rho(t) \simeq
\int_{1/\lambda^2}^{\infty} dk\ (\lambda k)P(k) e^{-t/\tau_{\lambda k}} \ .
\ee
Since it decays to zero eventually, the apparent active phase implied in Eq.~(\ref{qbeta}) is in fact an inactive one.
We remark that the inactive phase is different from the usual one in which
the density decays exponentially in time.
The density in Eq~(\ref{relax_dyn}) does not decay exponentially
fast but extremely slow due to the broad distribution of relaxation times.
For example, in SF networks with $P(k) \sim k^{-\gamma}$, one finds
\be\label{log_relax}
\rho(t) \sim [\ln t]^{-(\gamma-2)}~.
\ee

\begin{figure}
\includegraphics*[width=\columnwidth]{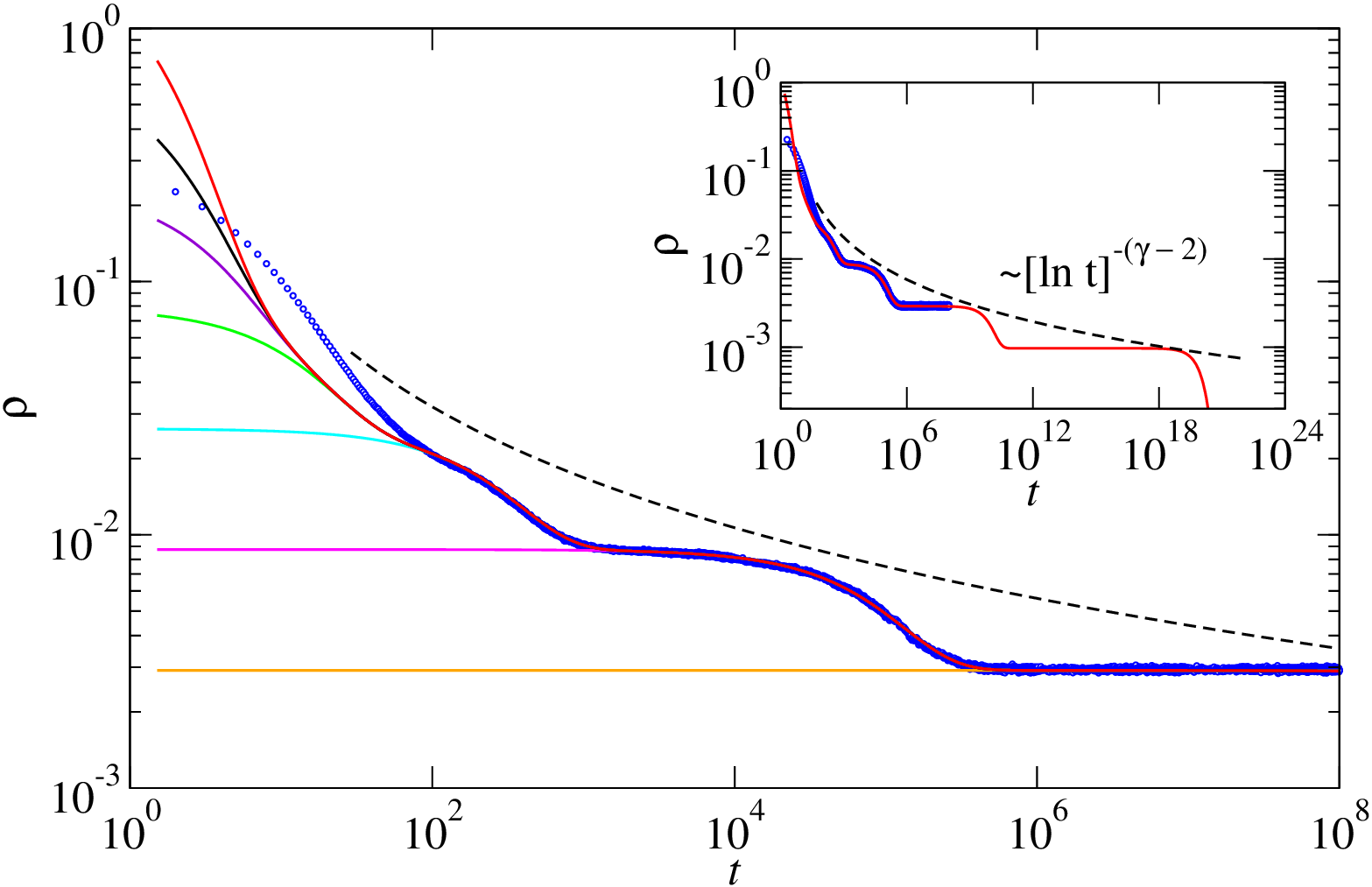}
\caption{(Color online) $\rho(t)$ in the $(3,3)$-flower of $G=8$.
The data points are the SIS model simulation result.
The solid curves are obtained from the direct integration of
Eq.~(\ref{relax_dyn}) with two fitting parameters $a_1=0.8$ and $a_2=0.325$
for $\tau_{\lambda k} = a_1 e^{a_2 \lambda k}$ and
with the truncated degree
distribution function; from bottom to top, $P(k)$ is truncated
by $k\geq k_{\rm max}/2^n$ with $n=1,2,..,7$.
Inset compares the long-time behavior predicted in our theory
(red solid curve)
with the currently available numerical data~(blue dots).}
\label{fig2}
\end{figure}

Equation~(\ref{relax_dyn}) is numerically tested in the $(3,3)$-flowers
at $\lambda=0.56<\lambda_c$,
which is shown in Fig.~\ref{fig2}.
It does not decay exponentially in time. Instead, the density is trapped
in a plateau for a
while and then decays to another plateau successively. Those plateaus are
the evidence for the existence of the metastable local active domains.
In the $(u,v)$-flower, degrees of nodes
are discretized as $k=2^n$ with
$n=1,2,..$.
So the size $V_k \sim
\lambda k$ of the local active domains and their life time $\tau_{\lambda k}
\sim e^{a \lambda k}$ are also discrete. This discreteness results in the
plateaus.
As shown, the data in Fig.~\ref{fig2} are fitted to Eq.~(\ref{relax_dyn})
very well for large $t\gtrsim 10^2$.
The dashed curve therein represents the overall decay given by
Eq.~(\ref{log_relax}).

The slow dynamics given in Eqs.~(\ref{relax_dyn}) and (\ref{log_relax})
is reminiscent of a relaxation dynamics in the Griffiths
phase~\cite{Vojta06a,Hooyberghs03,Vojta06,Munoz10}. In a disordered system,
disorder fluctuations may generate local domains which behave differently
from the bulk. Denoting the probability that such a domain of size $\xi$ is
realized as $P(\xi)$ and the relaxation time therein as $\tau(\xi)$, a
physical quantity $f$ relaxes to its stationary value $f_s$ as $\delta f(t)
= f(t) - f_s \sim \int d\xi P(\xi) e^{-t/\tau(\xi)}$. In the Griffiths
phase, the relaxation dynamics is dominated by rare events encoded in
the tail of $P(\xi)$ with long characteristic time scales
$\tau(\xi)$~\cite{Vojta06a}. In our case, the slow dynamics is originated
from the irreversible fluctuation near hubs.

The previous argument shows that the SIS model on the unclustered network
can be in the inactive Griffiths phase for
$\lambda_{\rm c}^{\rm QMF} < \lambda < \lambda_{\rm c}$
with a certain nonzero $\lambda_{\rm c}$.
It also provides a hint on the mechanism for the phase transition into
an active phase.
Inside the Griffiths phase, local active domains of size
$\sim \lambda k$ are separated~[unclustered network] and metastable.
As $\lambda$ increases, the size of local domains grow and they begin to
overlap each other. The active domains become globally stable when
they will form a percolating giant cluster at a certain
threshold value of $\lambda$.
That is to say, the epidemic transition is triggered by a percolation
transition of the local active domains.

Note that the local active domains nucleate around high degree nodes in the
degree-descending order. So the uncovered mechanism leads to the conjecture
that the unclustered network with $\lambda_c \neq 0$ should have a nonzero
percolation threshold in the DOP model.
Recall that the DOP is suggested as a percolation model where nodes are
occupied in the degree-descending order~\cite{comment}.
A nonzero percolation threshold $p_c$ for the node occupation probability
implies that high degree nodes are well separated from each other.
Hence, in the context of the SIS
model, one requires a nonzero value of $\lambda$ for the local active
domains~(of size $\sim \lambda k$) to form a percolating cluster.

We provide a numerical evidence of our claim.
Figure~\ref{fig1} (b) shows the the percolation order parameter $g$, the
density of nodes in the largest cluster, for the DOP in the same
$(3,3)$-flowers used in (a). The percolation threshold $p_{\rm
c}$ is clearly nonzero, which is consistent with the fact that
$\lambda_c\neq 0$ as shown in Fig.~\ref{fig1} (a).
As another example for the validity of the claim,
one may revisit the aforementioned $k_{\rm max}$-star graph case
with large $L$. In this example, one can easily find that $p_{\rm c} \neq 0$
supporting our claim.

We also consider the opposite case with $p_{\rm c}=0$. It is achieved
only when any finite fraction of occupied nodes
in the DOP process are connected to each other to
form a percolation
cluster. Those networks with $p_c=0$ will be called the {\em clustered}
network. In the context of the SIS model, the local active domains in
the clustered network form a percolating cluster even in the limit
$\lambda\to 0$. Therefore, we expect that the epidemic threshold is zero in
the clustered network. The steady state density of the infected nodes
can be expected to
scale as
\be
\rho \sim \int_{1/\lambda^2}^{\infty}dkP(k) \ .
\ee
In comparison with Eq.~(\ref{qbeta}), the factor $(\lambda k)$ is
missing because a stable active domain consists mainly of
high-degree nodes with $k>1/\lambda^2$.
In scale-free networks with $P(k) \sim k^{-\gamma}$,
one obtains $\rho \sim \lambda^{2\gamma-2}$.
Note that the order parameter exponent $\beta = 2\gamma-2$ is different
from $\beta_{\rm QMF}=1$ obtained from the simple QMF theory where only
the largest eigenmode was taken into account~\cite{Mieghem12}.

\begin{figure}
\includegraphics*[width=\columnwidth]{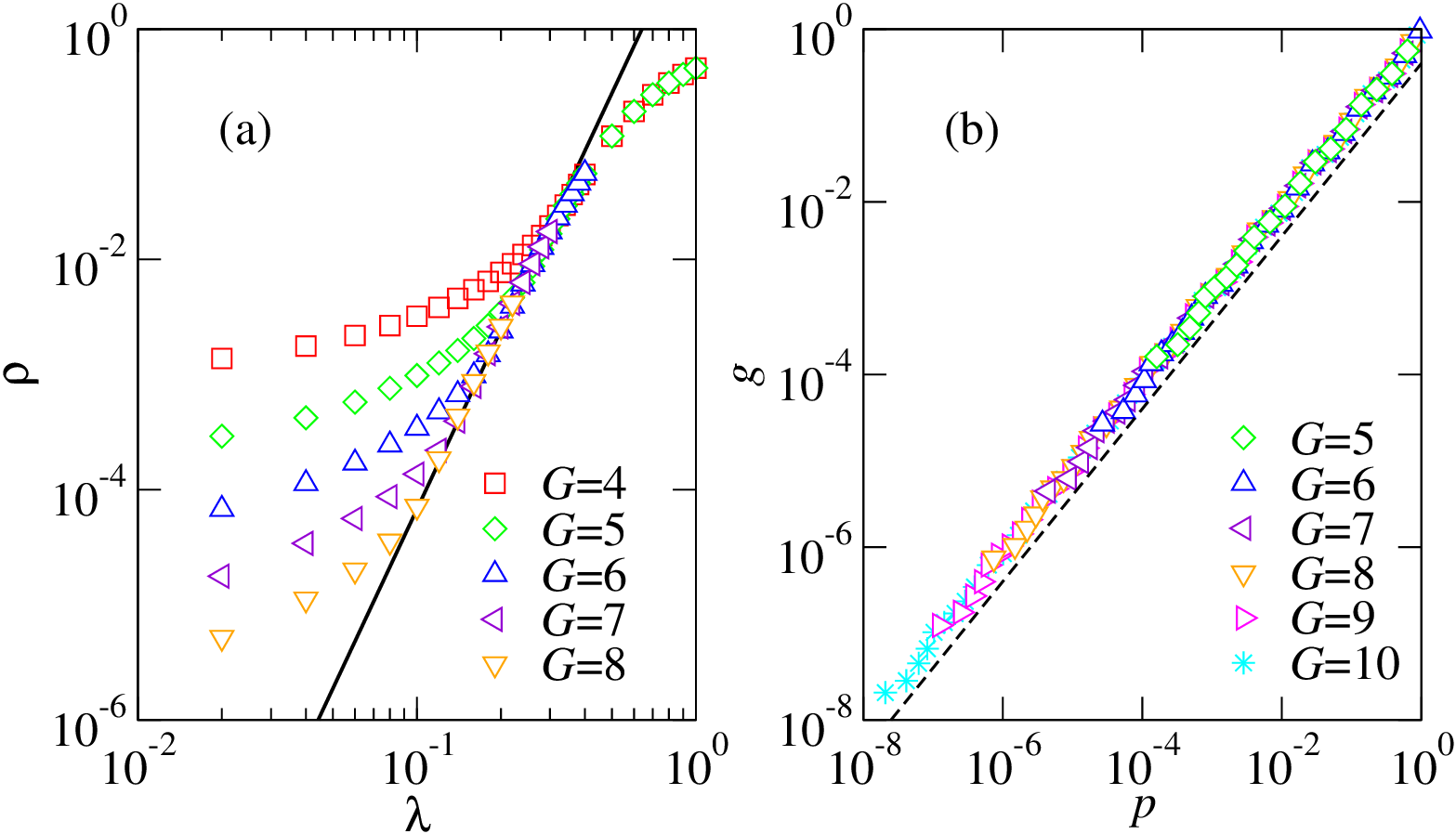}
\caption{(Color online) Density of infected nodes in the $(1,5)$-flower in (a). The largest cluster density in the DOP on the same flower in (b).
The solid lines in (a) has a slope $2\ln 6/\ln 2 \simeq 5.19$ while the dashed line in (b) has a slope 1.}
\label{fig3}
\end{figure}

A trivial example of the clustered network
is the $k_{\rm max}$-star graph with $L=1$. The density
of the largest cluster is given by $g=p$ implying that $p_c=0$.
Independently, $\lambda_{\rm c}=0$ obviously as discussed in
Ref.~\cite{Castellano10}.
A nontrivial example is the $(u,v)$-flowers
with $u=1$~(or $v=1$).
Recall that if two nodes are connected with an edge in a certain generation,
then they remain being connected afterward
when $u=1$~(or $v=1$).
So one can expect that high degree nodes are clustered.
In Fig.~\ref{fig3}(a), we present the SIS model simulation results on the
$(1,5)$-flowers at several generations $G$.
As $G$ increases, the data approach the theoretical prediction $\rho \sim
(\lambda-\lambda_c)^{\beta}$ with $\lambda_c=0$ and $\beta = 2\gamma-2 =
2\ln 6/\ln 2$. The DOP property is presented in
Fig.~\ref{fig3}(b). The scaling of $g \sim p$ therein indicates $p_c=0$.

What is the epidemic threshold in more interesting cases such as
generic random SF networks? Recently, extensive Monte Carlo simulations
were performed in random SF networks generated from the configuration
model~\cite{Ferreira12}.
However, due to a strong finite-size effect, it still remains
inconclusive whether
$\lambda_{\rm c}=0$ or not even with simulations of system sizes
up to $N=3\times 10^7$.

\begin{figure}
\includegraphics*[width=\columnwidth]{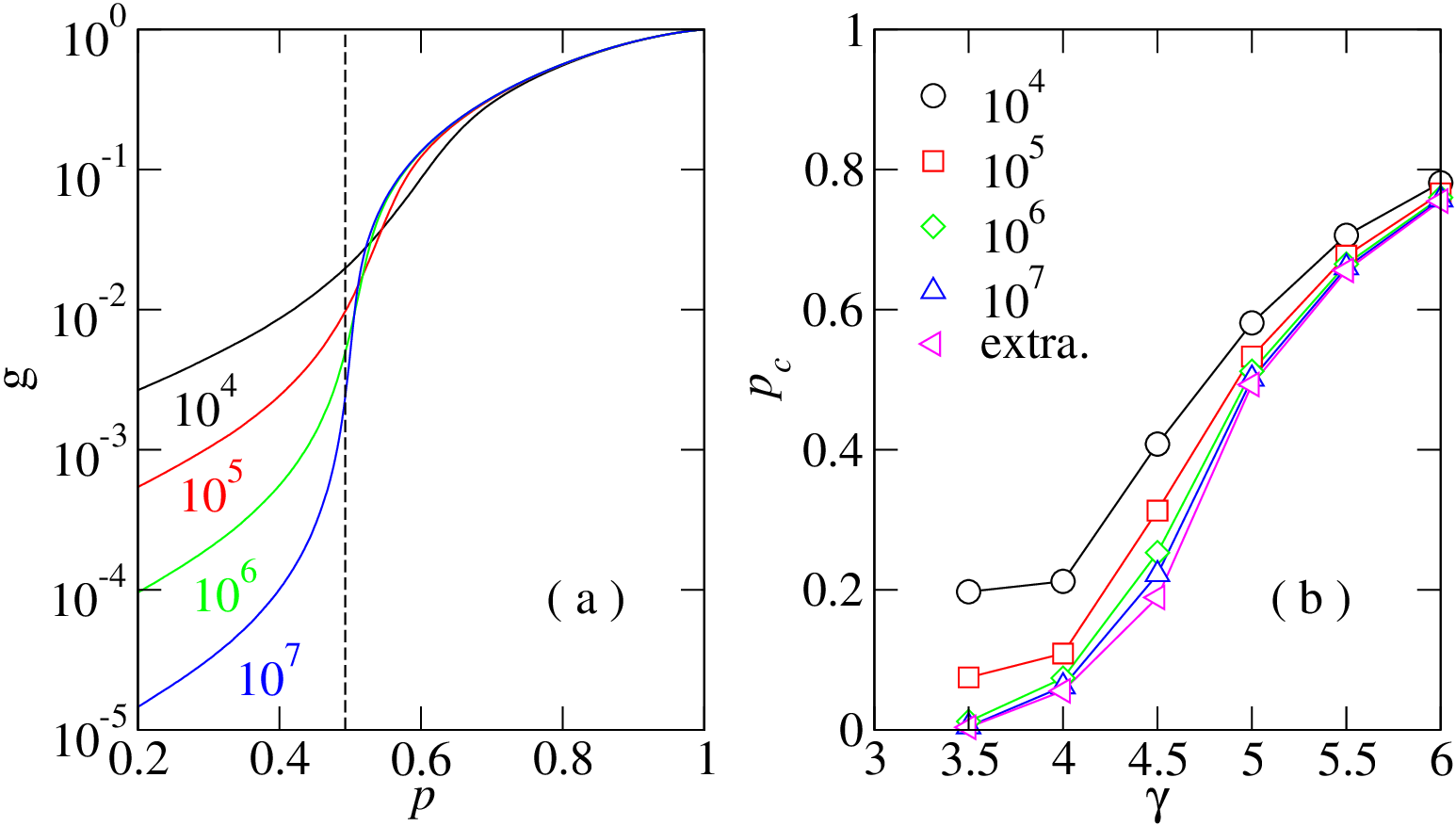}
\caption{(Color online) (a) The density of the largest DOP cluster in random SF networks
of the configuration model with $\gamma=5$. The dashed line denotes the
percolation transition point. (b) Maximum cluster
heterogeneity points and their extrapolated values. The
system sizes are $N=10^4,..,10^7$.}
\label{fig4}
\end{figure}

Alternatively we investigate
the DOP property of the configuration model networks.
In Fig.~\ref{fig4}(a) we present the percolation order parameter for the
network with $\gamma=5$.
As in the $(3,3)$-flower, the system undergoes the
percolation transition at a finite threshold. In order to estimate the
percolation threshold precisely, we make use of a so-called cluster
heterogeneity~(CH) which denotes the number of distinct cluster
sizes~\cite{Lee11}. It was shown~\cite{Noh11} that $p_{\rm c}(N)$ at which the
CH is maximum in networks of size $N$ converges to the
percolation threshold $p_{\rm c}$ in the infinite $N$ limit. In
Fig.~\ref{fig4}(b), we present the numerical data for $p_{\rm c}(N)$
at several values of $\gamma$ and their extrapolated values.
The percolation threshold is nonzero unless
a small $\gamma$ is considered.
Thus, the random SF networks
therein
belong to the unclustered network
class. It provides an indirect evidence that the epidemic threshold could
be nonzero on those SF networks.

In summary, we present a theoretical argument that the epidemic
threshold of the SIS model on complex networks is nonzero in the unclustered
network while it is zero in the clustered network.
This conclusion is drawn by taking account of the effect of
the irreversible fluctuation which was ignored in the QMF theory.
The fluctuation makes a local active domain unstable and leads to
the Griffiths phase.
Numerical simulations performed in the deterministic $(u,v)$-flowers support
our argument. We suggest that the clustering property of a network can be
determined by the DOP. By studying the DOP transition, the random SF
networks are shown to belong to the unclustered network unless a small degree exponent is considered. It suggests the
epidemic threshold in such SF networks is nonzero as opposed to the QMF
prediction.
Our work raises various interesting questions on the critical phenomenon
associated with the epidemic transition and the DOP transition, which are
left for future works.

This work was supported by the National Research Foundation of
Korea~(NRF) grant funded by the Korea government~(MEST)~(No. 2012-0005003).
We thank Prof. Hyunggyu Park for helpful discussions.


\begin{thebibliography}{99}
\bibitem{comment} The DOP is an extreme case of a percolation model with
a degree-dependent occupation probability considered in Ref.~\cite{Gallos05}.
\end{thebibliography}
\end{document}